\documentclass[aps,pra,amsmath,amssymb,amsfonts,superscriptaddress,twocolumn,footinbib]{revtex4-1}

\usepackage{graphicx}
\usepackage[english]{babel}
\usepackage{braket}
\usepackage{color}      
\usepackage[pdfpagelabels,plainpages=false]{hyperref}   

\newcommand{\Fig}[1]{Fig.~\ref{#1}}
\newcommand{\Sec}[1]{Sec.~\ref{#1}}

\newcommand{\rmd}{{\rm d}}
\newcommand{\rme}{{\rm e}}

\def\figpath{.}

\begin{document}

\title{Symmetry Breaking and the Geometry of Reduced Density Matrices}

 \author{V. \surname{Zauner}}
 \affiliation{Vienna Center for Quantum Technology, University of Vienna, Boltzmanngasse 5, 1090 Wien, Austria}
    \author{D. \surname{Draxler}}
 \affiliation{Vienna Center for Quantum Technology, University of Vienna, Boltzmanngasse 5, 1090 Wien, Austria}
  \author{L. \surname{Vanderstraeten}}
   \affiliation{Ghent University, Faculty of Physics, Krijgslaan 281, 9000 Gent, Belgium}
  \author{J. \surname{Haegeman}}
   \affiliation{Ghent University, Faculty of Physics, Krijgslaan 281, 9000 Gent, Belgium}
 \author{F. \surname{Verstraete}}
 \affiliation{Vienna Center for Quantum Technology, University of Vienna, Boltzmanngasse 5, 1090 Wien, Austria}
 \affiliation{Ghent University, Faculty of Physics, Krijgslaan 281, 9000 Gent, Belgium}

\begin{abstract}
The concept of symmetry breaking and the emergence of corresponding local order parameters constitute the pillars of modern day many body physics. We demonstrate that the existence of symmetry breaking is a consequence of the geometric structure of the convex set of reduced density matrices of all possible many body wavefunctions. The surfaces of these convex bodies exhibit non-analyticities, which signal the emergence of symmetry breaking and of an associated order parameter. We illustrate this with three paradigmatic examples of many body systems exhibiting symmetry breaking: the quantum Ising model, the classical Ising model in two dimensions at finite temperature and the ideal Bose gas in three dimensions at finite temperature. This quantum state based viewpoint on phase transitions provides a unique novel tool for studying exotic quantum phenomena.
\end{abstract}

\maketitle

Ground states of quantum many body Hamiltonians composed of local interactions are very special: in order to minimize the energy expectation value they have extremal local correlations, but those correlations must be compatible with the global symmetry of the many body Hamiltonian such as translation invariance. The competition between those two requirements is responsible for the emergence of the typical long-range properties as exhibited in strongly correlated materials. This is best illustrated by the $S=1/2$ Heisenberg antiferromagnet: the energy density would be minimized if all nearest neighbor reduced density matrices (RDMs) were singlets, but due to the monogamy properties of entanglement \cite{monogamy1,monogamy2}, this is not compatible with the translation invariance of the ground state. Hence the symmetry requirements smear the entanglement out into a globally entangled state with algebraically decaying correlations and (quasi) long range order.

\begin{figure*}
 \centering
 \includegraphics[width=\linewidth,keepaspectratio=true]{\figpath/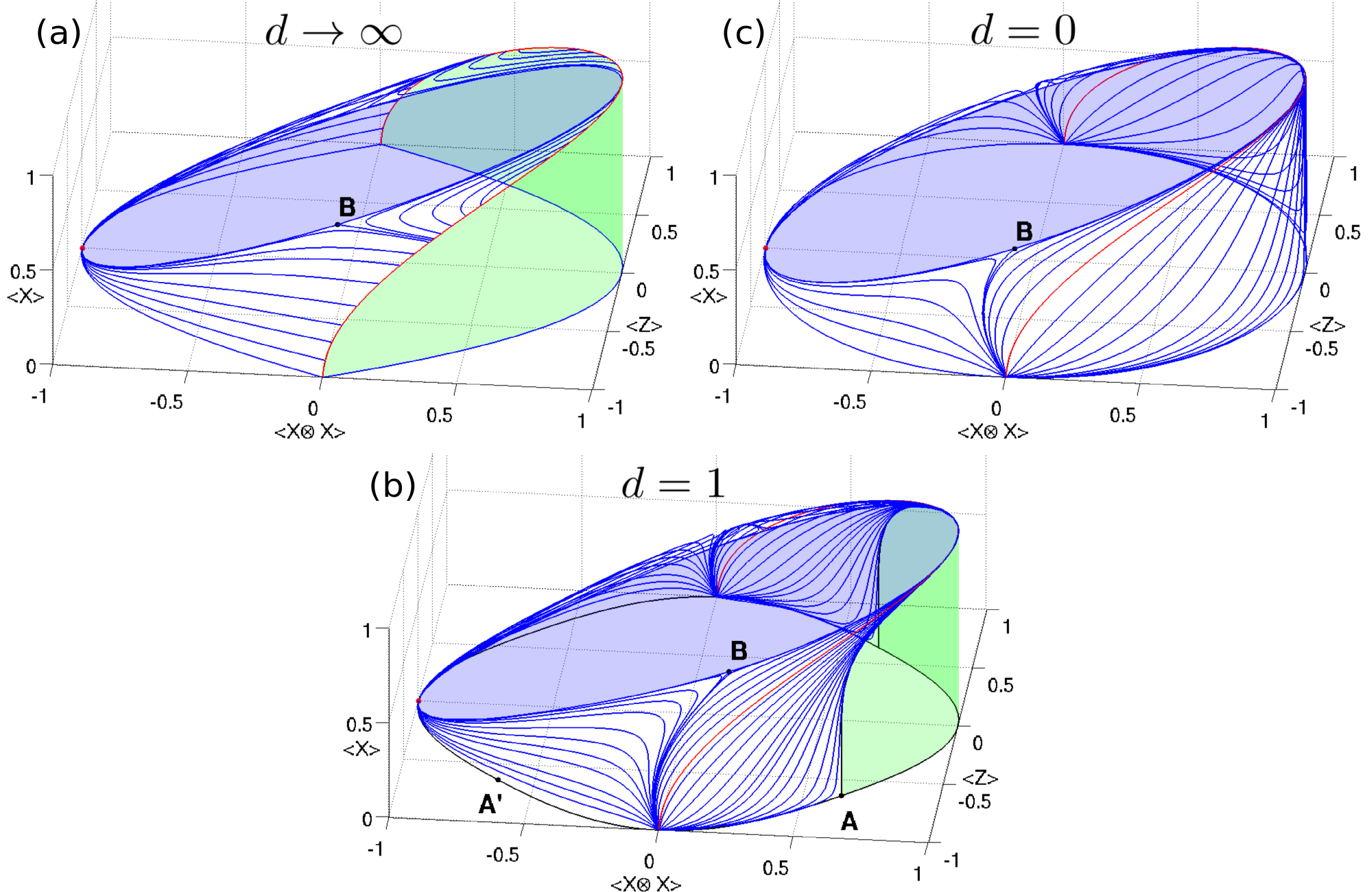}
 \caption{Convex sets of the nearest neighbor correlation $\braket{XX}$, transverse magnetization $\braket{Z}$ and longitudinal magnetization
$\braket{X}$ for all possible translation  invariant states on an infinite lattice of $S=1/2$ spins in \textbf{(a)} $d\to\infty$ and \textbf{(b)} $d=1$
spatial dimensions, as well as \textbf{(c)} $d=0$ spatial dimensions (all possible states of two spins). We plot the surfaces of extreme points of these sets,
corresponding to ground states of \eqref{eq:IsingHam}
for $J=\pm1$ and various values of $B_{z}$ and $B_{x}>0$
(due to symmetry we only plot the upper half). Blue lines represent points with constant $B_{x}$ and varying $B_{z}$.
In \textbf{(b)}, the black line corresponds to the exact solution for $B_{x}=0$ \cite{Pfeuty} and points \textbf{A} and \textbf{A$'$} mark critical points, where at \textbf{A} a
ruled surface (green) emerges, signaling a degeneracy of the ground state, which leads to symmetry breaking and a finite value of the order parameter $\braket{X}$,
with a critical exponent of $1/8$ (see main text). 
In all three cases, the red line corresponds to $J=0$ and thus separates regimes of
ferromagnetic ($J>0$) and antiferromagnetic ($J<0$) coupling. For further details, in particular about the blue top plane, 
see \cite{supp}.
}
\label{fig:fig1}
\end{figure*}

This competition is nicely reflected in the convex set of the RDMs of all possible pure and mixed quantum many body states of the entire system 
\footnote{When restricting oneself to RDMs of pure states, this problem is related to identifying the joint numerical range of a set of operators, which is not necessarily convex anymore \cite{numerical_range1,numerical_range2}.}. The relevance of such convex sets was first realized by Gibbs \cite{Gibbs,Maxwell,Israel}, as all thermodynamic properties of the systems under interest can be read off from the geometric features of this set and phase transitions correspond to non-analyticities, which arise by considering convex hulls of analytic functions \cite{VdW}. Such convex sets were also used in the context of the $N$-representability problem in quantum chemistry to illustrate quantum phase transitions in lattice systems \cite{N-representability1,N-representability2,N-representability3,N-representability4}, and in the context of discussions on the monogamy of entanglement and mean field theory in \cite{faithful}.

Along this line, let us for example consider a lattice of spin-$1/2$ quantum systems, take random quantum states $\rho_\alpha$, and  make a scatter plot of the expectation values $\braket{Z}\equiv{\rm Tr}\left(\rho_\alpha\sum_i Z_i\right)$ and $\braket{XX}\equiv{\rm Tr}\left(\rho_\alpha\sum_{\braket{ij}}X_iX_j\right)$, where $X$ and $Z$ are Pauli matrices and the sum is over nearest neighbors only. This is equivalent to restricting to translation invariant states $\rho^{\rm TI}_{\alpha}$ and measuring the 1 or 2-site observables $Z$ and $X\otimes X$, for which only the 2-site RDM is needed. As these terms do not commute, a large expectation value  $\langle XX\rangle$ will lead to a small expectation value  $\langle Z\rangle$,
giving rise to a curved boundary of the generated set. Due to the convexity of the set of 2-site RDMs, the generated body is also convex and corresponds to a two-dimensional projection of the full 15-dimensional set of all possible 2-site RDMs. The extreme points of this set correspond to ground states of a family of quantum Ising Hamiltonians of the form
\begin{equation}
\mathcal{H}=-J\sum_{\braket{i,j}}X_iX_j - B_{z}\sum_i Z_i.
\label{eq:Ising0}
\end{equation}
Indeed, surfaces of constant energy are represented by lines $E = -J\braket{XX} - B_{z}\braket{Z}$ in this plot, where their orientation is given by the parameters $J$, $B_{z}$ and their distance to the origin is proportional to (minus) the energy. Hence the expectation values of the states with minimal energy must correspond to extreme points for which the lines are tangent to the convex set, or equivalently, every point on the boundary of the generated set corresponds to the ground state of \eqref{eq:Ising0} with parameters given by the orientation of the tangent line through that point.

The situation becomes much more interesting and the presence of symmetry breaking becomes immediately evident when adding an extra axis corresponding to the expectation value of $\braket{X}\equiv{\rm Tr}\left(\rho_\alpha\sum_i X_i\right)$ to the scatter plot. The extreme points of the resulting convex set now correspond to ground states of the quantum Ising model including a symmetry breaking longitudinal field
\begin{equation}
 \mathcal{H}=-J\sum_{\braket{i,j}}X_iX_j - B_{z}\sum_i Z_i - B_{x}\sum_{i}X_{i}.
 \label{eq:IsingHam}
\end{equation}
In \Fig{fig:fig1} we show the surface of this set in zero, one and infinite spatial dimensions (see \cite{supp} for a scatter plot).

For an infinite system in $d\geq 1$ spatial dimensions we witness the emergence of a ruled surface with all lines parallel to the new axis, which turns out to be the defining signature for symmetry breaking.
Indeed, all points on such a line are ground states to the same instance of \eqref{eq:IsingHam} -- with parameters given by the orientation of the tangent plane
\footnote{Here, all tangent planes with normal vectors $\vec{n}=\left(1,B_{z},0\right)$ with $\left|B_{z}\right|\leq 1$ will touch the set on a line of the ruled surface, instead of a single point.} -- but with different values of $\braket{X}$. This implies that the ground state is not unique and there is symmetry breaking, as
an infinitesimal perturbation of the form of a longitudinal magnetic field $\epsilon \sum_i X_i$ to the Hamiltonian \eqref{eq:Ising0} will break the symmetry, and make sure that the magnetization of the ground state will be polarized in the $x$-direction with a magnitude given by the extreme points of the convex set lying on the border of the ruled surface. $\braket{X}$ is then obviously the order parameter, and the shape of the border of the ruled surface encodes all the information about the ground state expectation values such as the order parameter as a function of the (transverse) magnetic field $B_{z}$.

Furthermore, from \Fig{fig:fig1} we observe three additional remarkable but obvious facts. (i) The convex set of the zero-dimensional case completely contains the one-dimensional set, which in turn completely contains the infinite-dimensional case. This reflects the fact that more and more symmetry constraints (e.g. translation invariance along an increasing number of spatial dimensions) restrict the convex set of possible 2-site RDMs further and further.
(ii) The ruled surfaces only arise in the thermodynamic limit, hence demonstrating the need for the well known fact that the order parameter is obtained by first taking the limit of the system size to infinity and only then the longitudinal magnetic field to zero. In \cite{N-representability3,N-representability4}, the concept of speed was introduced to describe the curvature of convex sets of systems on finite lattices, and observed to diverge when doing finite size scaling.
(iii) We can extract critical exponents by investigating the geometry of the convex set around the critical point 
\footnote{In the one-dimensional case, the dependence of $\braket{X}$ on the magnetic field $B_{z}$ in the symmetry broken phase slightly below the critical point $B_{z}/J=1-\varepsilon$ with $\varepsilon\ll1$ can be recovered from the orientation of the tangent plane $E = -J\braket{XX} - h\braket{Z}$ as $\frac{\rmd \braket{XX}}{\rmd \braket{Z}}=\varepsilon-1$ to find that indeed in this regime
$\braket{X}\propto\Big(\frac{\rmd \braket{XX}}{\rmd \braket{Z}}+1\Big)^{1/8}$.}.
More generally, any thermodynamic property of the system such as susceptibilities can be read off from this convex set and the properties of its surface, hence demonstrating the power of such convex set plots (see also \cite{supp}).


\begin{figure}[t]
\centering
 \includegraphics[width=\linewidth,keepaspectratio=true]{\figpath/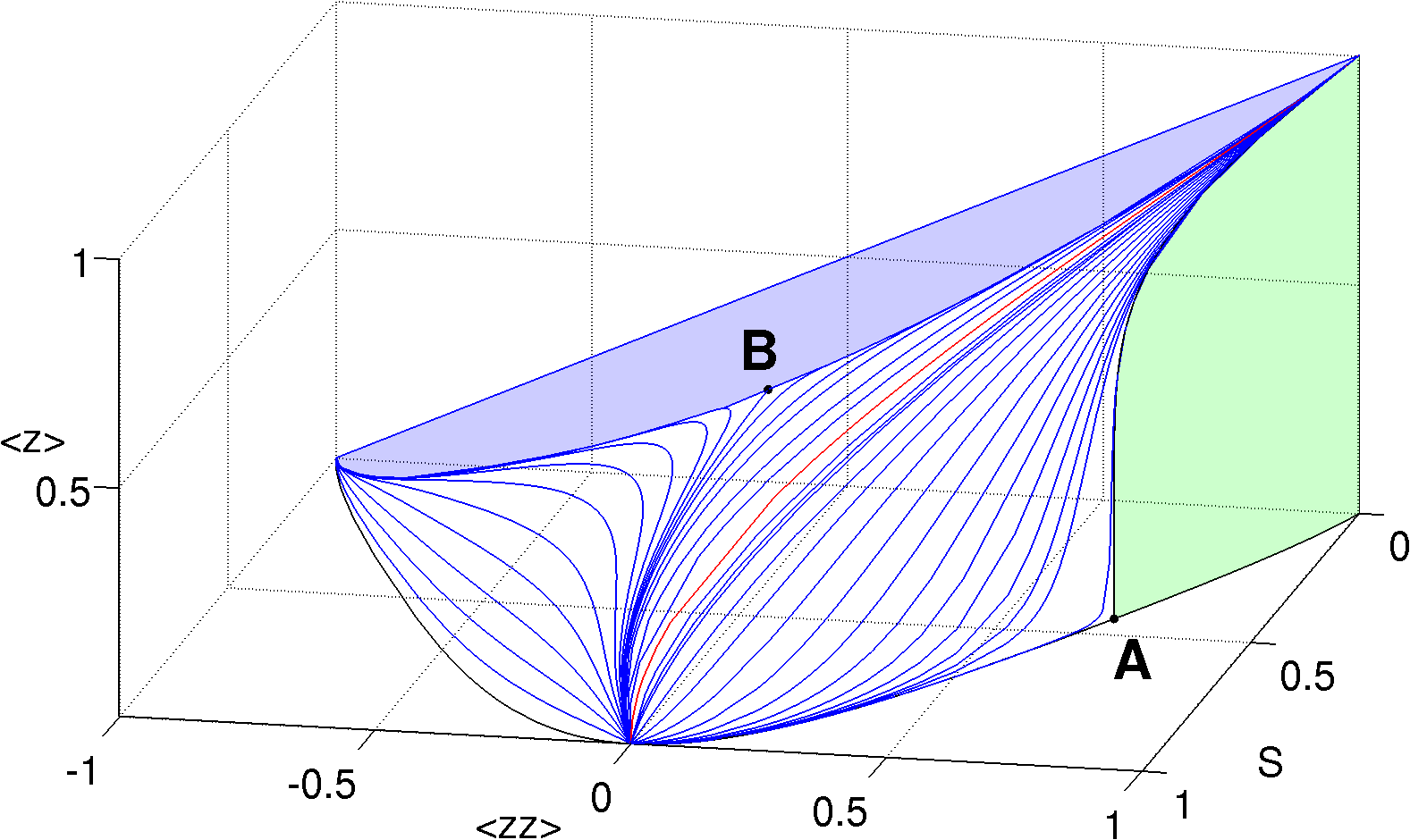}
\caption{
 Convex set of the average nearest neighbor correlation $\braket{z_iz_j}$, the entropy per site $S$, and the expectation value of the magnetization per site $\braket{z_i}$ for all possible probability distributions of classical spin configurations on an infinite 2D square lattice. The surface of extreme points corresponds to Gibbs states of the classical Ising model on a square lattice, given by $E = -J \sum_{<i,j>} z_iz_j -h \sum_i z_i$, for a range of values of the interaction strength $J$, the magnetic field $h$ and temperature $T$ \cite{supp}. The blue lines correspond to points with constant magnetic field $h$ and varying temperature $T$, while the red line corresponds to $J=0$, separating the ferromagnetic ($J>0$) from the antiferromagnetic ($J<0$) regime. 
 The black line represents the exact solution at $h=0$ \cite{Onsager}. Beyond the critical point \textbf{A} an emerging ruled surface (green) again signals symmetry breaking. The bifurcation point \textbf{B} with parameters $J=-1$ and $h=4$ gives rise to an exponentially degenerate lowest-energy state with a non-zero value of the entropy as $T\to0$.
 }
\label{fig:fig2}
\end{figure}

\begin{figure}[t]
\centering
\includegraphics[width=\linewidth,keepaspectratio=true]{\figpath/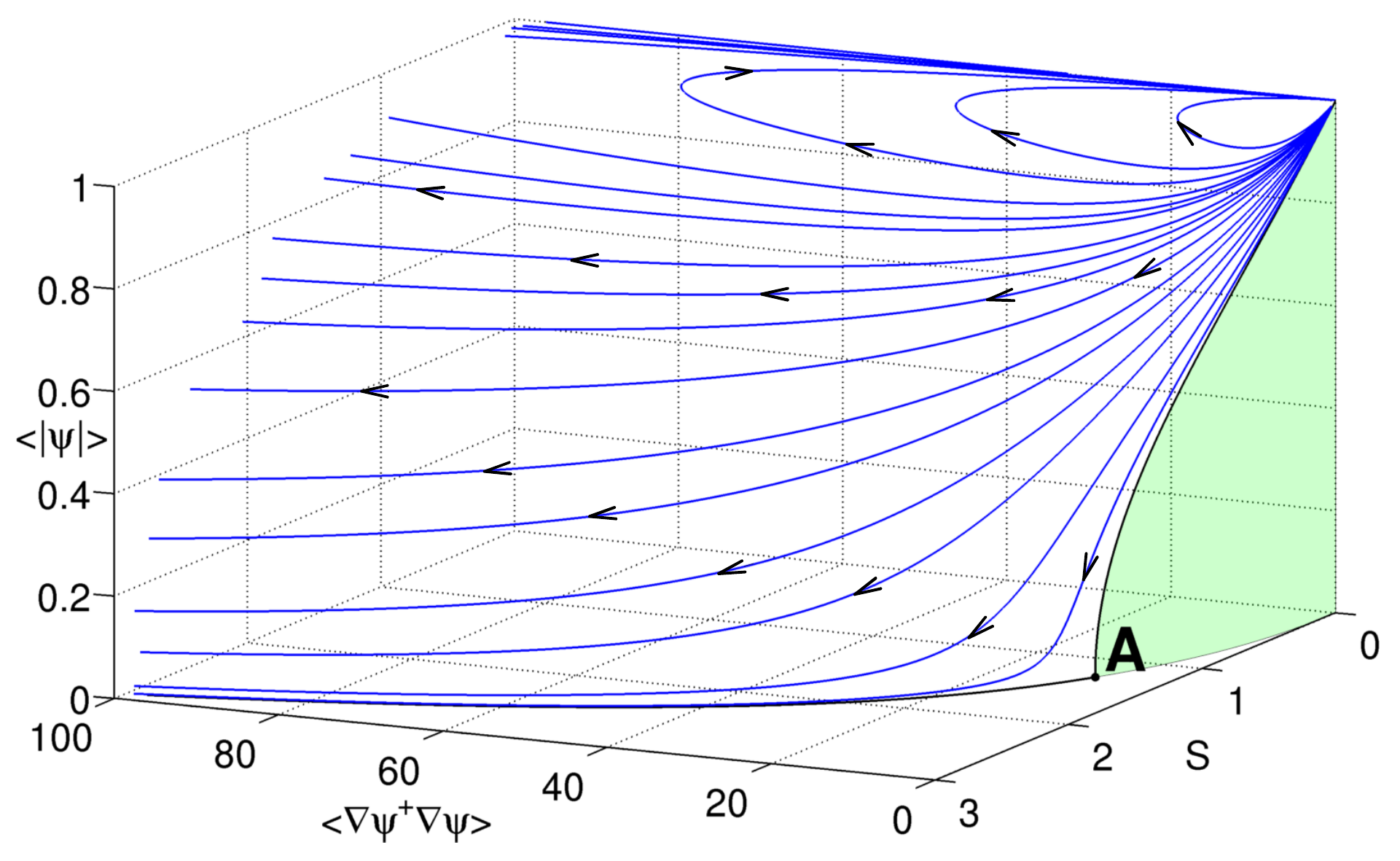}
\caption{
Convex set of the kinetic energy $\braket{\nabla\psi^{\dagger}\nabla\psi}$, the entropy $S$ and (the absolute values of) the order parameter $\braket{\psi}$ for all possible states of free bosons in three dimensions with fixed density. We show the surface of this set corresponding to Gibbs states of the ideal Bose gas
for various values of the temperature $T$ and $v\in\mathbb{R}$.
The blue lines correspond to RG flows through parameter space as indicated by the black arrows while the black line corresponds to $v=0$ and displays a critical point \textbf{A} with critical exponent of $1/2$ at a critical temperature $T_{c}$ \cite{supp}. Beyond this point a ruled surface emerges, witnessing symmetry breaking and the occurrence of Bose-Einstein condensation.
}
\label{fig:fig3}
\end{figure}

Notice that in these pictures ruled surfaces and non-analyticities arise without any reference to any underlying Hamiltonian. It is rather the choice of a collection of plotted observables that enables access to thermodynamic properties of a corresponding Hamiltonian defined by this collection. The occurrence of symmetry breaking is therefore encoded in the geometrical structure of a certain projection of the convex set of all possible RDMs, and quantum phase transitions are a consequence of the geometry of such convex sets.

In the case where one would like to learn about symmetry breaking phase transitions of a particular Hamiltonian without knowing the order parameter, one can still use a random observable as additional axis. In general a random observable will have a finite contribution of the true order parameter, thus still generating ruled surfaces, albeit not with maximal extent. In order to find the true order parameter one can then optimize over random observables to obtain the maximal extent of the emerged ruled surfaces \cite{supp}.


The general features of the plots in \Fig{fig:fig1} are clearly generic for second order phase transitions. Note that in the case of first order phase transitions, one would not need to add an extra axis (corresponding to the order parameter) to witness symmetry breaking, and this would already be present at finite size (see \cite{faithful} for an example).

A natural question is whether a similar picture emerges in the case of classical statistical physics. As opposed to the non-commutativity of the terms in a quantum Hamiltonian, classical phase transitions emerge as a consequence of the competition between the internal energy $E$ and the entropy $S$ in the free energy $F=E-TS$. As the free energy is a linear functional of energy and entropy, we expect similar convex sets as for the quantum case when we make a scatter plot with respect to all possible probability distributions of the expectation values of energy, entropy and the order parameter (see \Fig{fig:fig2} for the classical two-dimensional Ising model). Remarkably, we obtain a very similar picture as for the quantum case. The extreme points of the convex set now correspond to expectation values for Gibbs states. 
Note also that pictures involving temperature and/or pressure on the axes would not make sense in this setting, as those quantities are not defined for general probability distributions out of equilibrium.

As a final example, let us consider a 3D quantum system of free bosons in the continuum at finite temperature where we expect to find Bose-Einstein condensation.
Motivated by the above findings we plot the expectation value of the kinetic energy, the entropy, and a $U(1)$ symmetry breaking order parameter $\braket{\psi}$, with respect to all possible wavefunctions to obtain a meaningful convex structure (see \Fig{fig:fig3}). The extreme points of this set correspond to Gibbs states of the Hamiltonian $H = \int_{V} {\rm d^3}x\, \frac{1}{2m}\nabla\psi^{\dagger}(x)\nabla\psi(x)-v[\psi(x) + \psi^{\dagger}(x)]$ at fixed density $\rho=1$, where again a symmetry breaking term has explicitly been added. A ruled surface emerging below the critical temperature beautifully signals the onset of Bose-Einstein condensation, where the equilibrium state is not unique and can be parameterized by a finite value of 
$\braket{\psi}$. Again, the critical exponent can be extracted from the change of the orientation of the tangent plane around the critical point.

In conclusion, we investigated the convex structure of reduced density matrices of quantum many body systems and marginal probability distributions of models from statistical mechanics and illustrated how the concept of symmetry breaking emerges very naturally through the appearance of ruled surfaces at the boundaries of these sets. As these sets exist without any prior notion of an underlying Hamiltonian, this shows that the occurrence of symmetry breaking is encoded in the geometrical structures of the convex sets of all possible RDMs or marginal probability distributions. This picture seems to capture all the thermodynamically relevant features of many body systems in an extremely concise way. It would therefore be very interesting to classify all possible ruled surfaces that can arise on such convex sets.

Our work is very close in spirit to the original groundbreaking papers of Gibbs 
\cite{Gibbs,Israel}, which clarified that phase transitions and the coexistence of different phases can be understood in terms of non-analyticities in the parametrization of the surface of thermodynamic diagrams, and to ideas developed in the context of N-representability \cite{N-representability1,N-representability2,N-representability3,N-representability4} for describing quantum phase transitions in fermionic systems. It provides an explicit construction of the famous thermodynamic surface of Maxwell \cite{Maxwell} for the case of classical and quantum spin systems, and illustrates very concisely the mathematical physics point of view of symmetry breaking as a breakdown of ergodicity \cite{Fannes}. It also complements the ideas developed in \cite{RDM_order_parameter1,RDM_order_parameter2}, where a systematic procedure for finding order parameters was developed by contrasting the RDMs of the low-lying excited states in finite size quantum many body systems. Note however that the starting point of our work is very different: we make no a priori reference to a Hamiltonian, and just make a scatter plot with respect to all possible many body wavefunctions and/or probability distributions. Only the choice of observables relates the obtained convex set to the ground/equilibrium state properties of a corresponding Hamiltonian. Finally, the works \cite{geometry,Zyck} reported on the convex structure of expectation values of separable density matrices; in retrospect, those are the convex sets obtained in the mean-field regime of an infinite-dimensional lattice, as illustrated in \Fig{fig:fig1}(a).

One remaining open question is how topological phase transitions in the ground states of two- and higher-dimensional quantum systems fit in this description, as these cannot be characterized in terms of local order parameters. The tensor network description of quantum states might yield one possible resolution, as the topological order induces certain symmetries onto the virtual boundary theory of the tensor network \cite{mpoinjectivity1,mpoinjectivity2,mpoinjectivity3}. Topological phase transitions then correspond to symmetry-breaking phase transitions in the virtual boundary theory \cite{boundaryham}, i.e. in the structure of the fixed-point subspace of the transfer matrix of the tensor network. These transitions can thus be characterized in terms of a local order parameter at the virtual level of the tensor network \cite{shadows}. By bringing this virtual operator back to the physical level, it can be associated to the non-local string order parameters that characterize the topological phase \cite{nonlocalorder}. When considering systems on a torus, the natural approach would hence be to plot the expectation value of such a Wilson loop around the torus; the different ground states in the topological phase can then be distinguished by different values of this (nonextensive and nonlocal) order parameter, and hence a ruled surface should emerge at the topological phase transition.

\begin{acknowledgments}
We thank J.I. Cirac, C. Dellago, W. De Roeck, M. Mari{\"e}n and D. Nagaj for inspiring discussions. This work was supported by the Austrian Science Fund (FWF): F4104 SFB ViCoM and F4014 SFB FoQuS, ERA Chemistry and the EC through grants QUTE and SIQS.
\end{acknowledgments}

\appendix


\section{Numerical Data}
\label{s:numerics}
In this section we give details on how the surfaces of the convex sets shown in Figures \ref{fig:fig1}-\ref{fig:fig3} of the main text were numerically obtained. For an approximation of the surface for quantum spin systems by drawing scatter plots of a large number of random states, see \Sec{s:scatter}.
\paragraph{Quantum spin-1/2 lattices.} In \Fig{fig:fig1} of the main text, the surface of set \textbf{(a)} for $d\to\infty$ was numerically obtained using semidefinite programming: as a consequence of the monogamy properties of entanglement \cite{monogamy1,monogamy2} and the quantum de Finetti theorem \cite{de_Finetti}, this set is equivalent to the convex set generated by all separable states \cite{faithful,geometry,Zyck}. For the particular case
of two $S=1/2$ spins, separability is completely determined by semidefinite constraints \cite{pt1,pt2} and the surface of the set can be obtained by minimizing the energy $E = -J\braket{XX} - B_{z}\braket{Z} - B_{x}\braket{X}$ with respect to all separable density matrices of two spins.

Set \textbf{(b)} for $d=1$ was obtained by doing extensive variational
matrix product ground state calculations \cite{TDVP}, while set \textbf{(c)} for $d=0$ was obtained by exact diagonalization of a system of 2 spins.

\paragraph{Classical Ising model in 2D on a square lattice.}
In \Fig{fig:fig2}, both the free energy per site $F$, local expectation values and the internal energy per site $E$ of Gibbs states were computed using tensor network state techniques for transfer matrix renormalization in models of classical statistical mechanics \cite{ClassTN1,ClassTN2}, from which the entropy per site was then evaluated as $S =(E-F)/T$.

\paragraph{Ideal Bose Gas in 3D at finite Temperature.}
The system of an ideal Bose gas in the presence of a $U(1)$-symmetry breaking term can be solved analytically \cite{Gunton} and
the thermodynamic extensive quantities plotted in \Fig{fig:fig3} are readily found  to be $\langle \psi\rangle = -\frac{v}{\mu}$, $ S = \frac{5}{2}\lambda^{-3}F_{5/2}(-\beta\mu) - \beta\mu\lambda^{-3}F_{3/2}(-\beta\mu)$ and $E_{\rm{kin}} = \frac{3}{2\beta^{2}\lambda^{3}}F_{5/2}(-\beta\mu)$ with  $S$ being the entropy and
the chemical potential $\mu$ always chosen such that $\rho = \frac{v^{2}}{\mu^{2}} + \lambda^{-3}F_{3/2}(-\beta\mu) = 1$. Further we have defined $\lambda^{2}=\frac{2\pi}{mT}$ and $F_{\sigma}(x) = \sum_{n=1}^{\infty}n^{-\sigma}\rme^{-nx} \; \forall x \geq0$. At $v=0$ the critical temperature is given by $T_{c}(\rho) = \frac{2\pi}{m\lambda_{c}^{2}}$ with $\lambda^{3}_{c}=\rho^{-1}F_{3/2}(0)$. 

The blue lines in \Fig{fig:fig3} correspond to momentum-shell RG-flows through parameter space, which can be performed exactly \cite{Singh}. For fixed particle density the RG flow equations are found to be $\beta(s) = \beta(0)\,\rme^{-2s}$ and $v(s) = v(0)\,\rme^{\frac{7}{2}s}$ for $s\in(-\infty,\infty)$.

\section{Additional Information on Fig. 1}
\label{s:fig1}
In this section we give some additional informations about the convex sets for 2-site spin-1/2 RDMs shown in \Fig{fig:fig1} of the main text. As mentioned there, in all 3 subplots the red line corresponds to $J=0$, representing separable ground states, and thus divides regimes of ferromagnetic ($J>0$) and antiferromagnetic ($J<0$) coupling. Especially in \textbf{(a)} this line also marks the boundary of the green ruled surface, which implies that there is symmetry breaking for all values of the fields at $J>0$ in $d\to\infty$ spatial dimensions. 

In set \textbf{(b)}, the corresponding order parameter for the critical point \textbf{A$'$} is the \textit{staggered} magnetization $\braket{(-1)^{i}X_{i}}$
and we would therefore need to add another axis corresponding to that order parameter to observe the corresponding ruled surface.

In all three sets, point \textbf{B} marks the endpoint of a bifurcation line corresponding to $J=-1$, $B_{x}=2$ and $B_{z}\to0^{+}$, which leads up to a top (blue) plane. 
Especially for $d=1$ the corresponding Hamiltonian 
\begin{equation}
 \mathcal{H}_{\rm TP}=\sum_{j}X_{j}X_{j+1} - B_{x}\sum_{j}X_{j}
\end{equation} 
is in fact classical and all eigenstates are product states in the $x$-basis. One can easily see however that the ground state is exponentially degenerate with growing system size and the degeneracy is given by the Fibonacci sequence $F_{n+1}$ where $n$ is the number of lattice sites. To determine the edge of this plane we consider an infinitesimal perturbation $\mathcal{H}_{1}=\alpha\sum_{j}X_{j} + \beta\sum_{j}Z_{j}$ away from this point, with $\alpha, \beta\ll 1$ and project this perturbation onto the Fibonacci subspace  
\begin{equation}
\mathcal{H}_{p}=\sum_i \left[1 +X_{i-1}\right]\left[\alpha Z_{i} + \beta \left(X_{i} -1\right)\right]\left[1 + X_{i+1}\right].
\end{equation} 
The points on the edge of the top blue plane then correspond to ground states of this projected Hamiltonian in the case of $d=1$ spatial dimensions. 
In other words, for all values of $\alpha,\beta$ we seek linear combinations of states within the degenerate  ground state subspace which maximize the magnetization along the direction $(\alpha,0,\beta)$.

This plane has the same orientation but slightly different boundaries in all three cases. Note that the surface of \textbf{(b)} lies in between the surfaces of the two extremal cases \textbf{(a)} and \textbf{(c)}, which again reflects the further restrictions on the convex sets of all possible RDMs imposed by additional symmetry constraints.

Any set obtained for a \textit{finite} one-dimensional chain of $N$ spins would look similar to \textbf{(c)}. With increasing $N$ the surfaces of these sets will be gradually deformed to
asymptotically yield the surface of set \textbf{(b)} as $N\to\infty$, i.e. only in this limit will the green ruled surface and thus symmetry breaking emerge.

\begin{figure}[t]
 \centering
 \includegraphics[width=0.95\linewidth,keepaspectratio=true]{\figpath/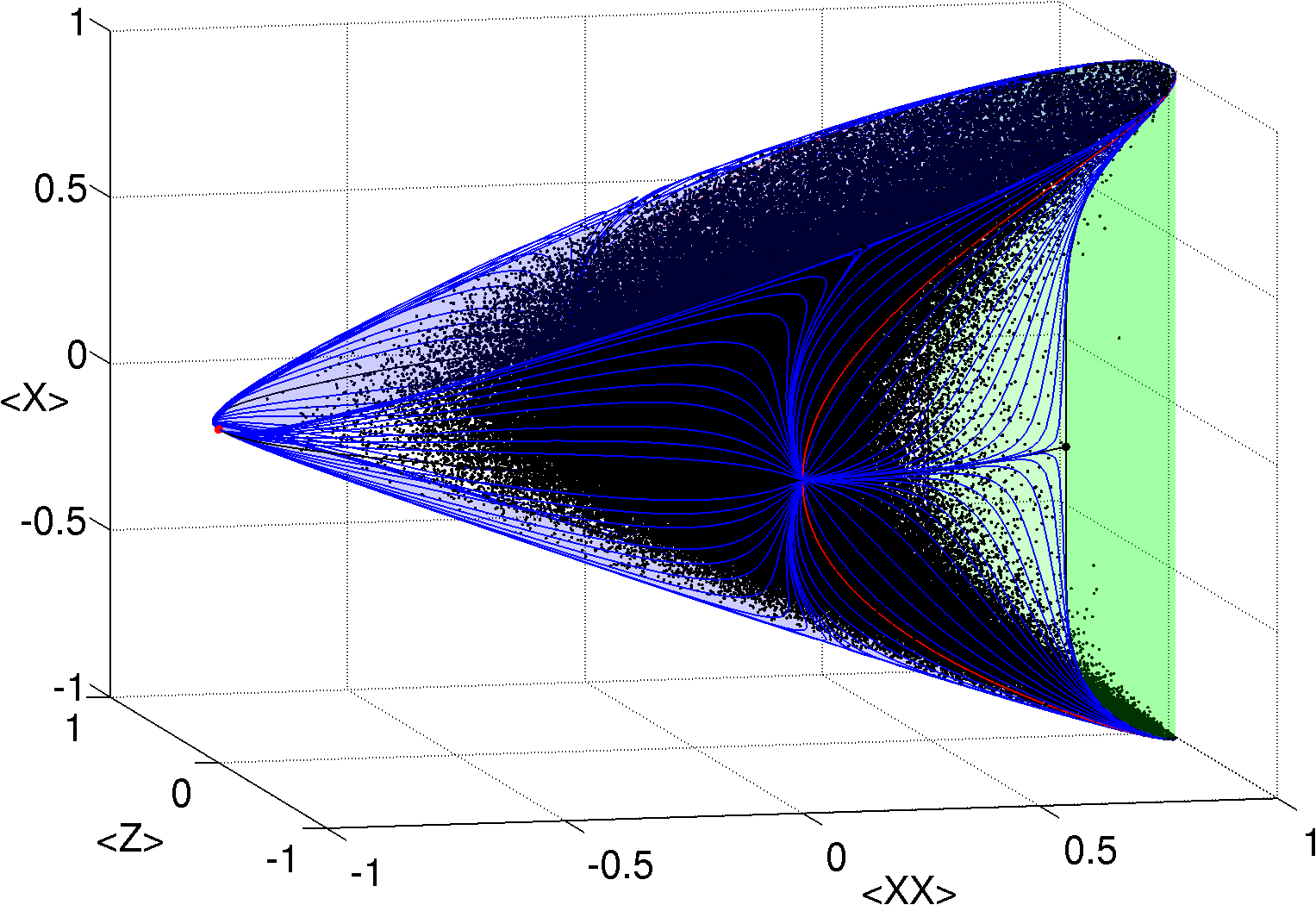}
 \caption{Scatter plot of observables of random states on a lattice of $S=1/2$ spins in $d=1$ dimensions, together with the surface shown in \Fig{fig:fig1}(b) of the main text. Here we explicitly plot both upper and lower half of the convex set for the sake of completeness. The black dots correspond to observables of single random quantum states with moderate entanglement. It is clearly visible how the green ruled surface emerges as two distinct lobes in the scatter plot of random states.}
 \label{fig:SFig1}
\end{figure}

\section{Scatter Plot for one-dimensional Quantum spin-1/2 lattices}
\label{s:scatter}
In \Fig{fig:SFig1} we show a scatter plot of observables of random states on a lattice of $S=1/2$ spins in $d=1$ dimensions, together with the surface already shown in \Fig{fig:fig1}(b) in the main text. The points in the plot were obtained by generating a large amount of random infinite matrix product states \cite{MPS} of low matrix dimension $D=2-10$, drawn from the unitary ensemble \cite{RMPS} and measuring the corresponding observables. The plot gives a beautiful practical demonstration of the procedure outlined on the first page of the main text. There it is argued that a scatter plot generated from drawing random states and calculating expectation values of local observables from (in this case) the 2-site RDM produces a convex set, whose surface is given by points corresponding to ground states of a family of Hamiltonians defined by the chosen collection of observables (in this case Eq. (2) from the main text). For a given set of Hamiltonian parameters, any points inside the set correspond to 2-site RDMs from (superpositions or mixtures of) excited states with respect to that Hamiltonian.

The surface shown in \Fig{fig:fig1}(b) of the main text is asymptotically obtained by taking the convex hull of the data points generated from a larger and larger amount of random states with varying bond dimension. \Fig{fig:SFig1} however shows that moderate bond dimensions of $D=2-10$ already yield a very good approximation of the true surface. The scatter plot also beautifully shows how the ruled surface emerges from the cloud of random data points in the form of two distinct lobes in the vicinity of the green ruled surface with hardly any points in between, whereas away from the the ruled surface the data points are distributed fairly homogeneously. We observe that this is the characteristic geometric phenomenon for the occurrence of symmetry breaking in terms of a random scatter plot.

\begin{figure*}[ht!]
 \centering
 \includegraphics[width=\linewidth,keepaspectratio=true]{\figpath/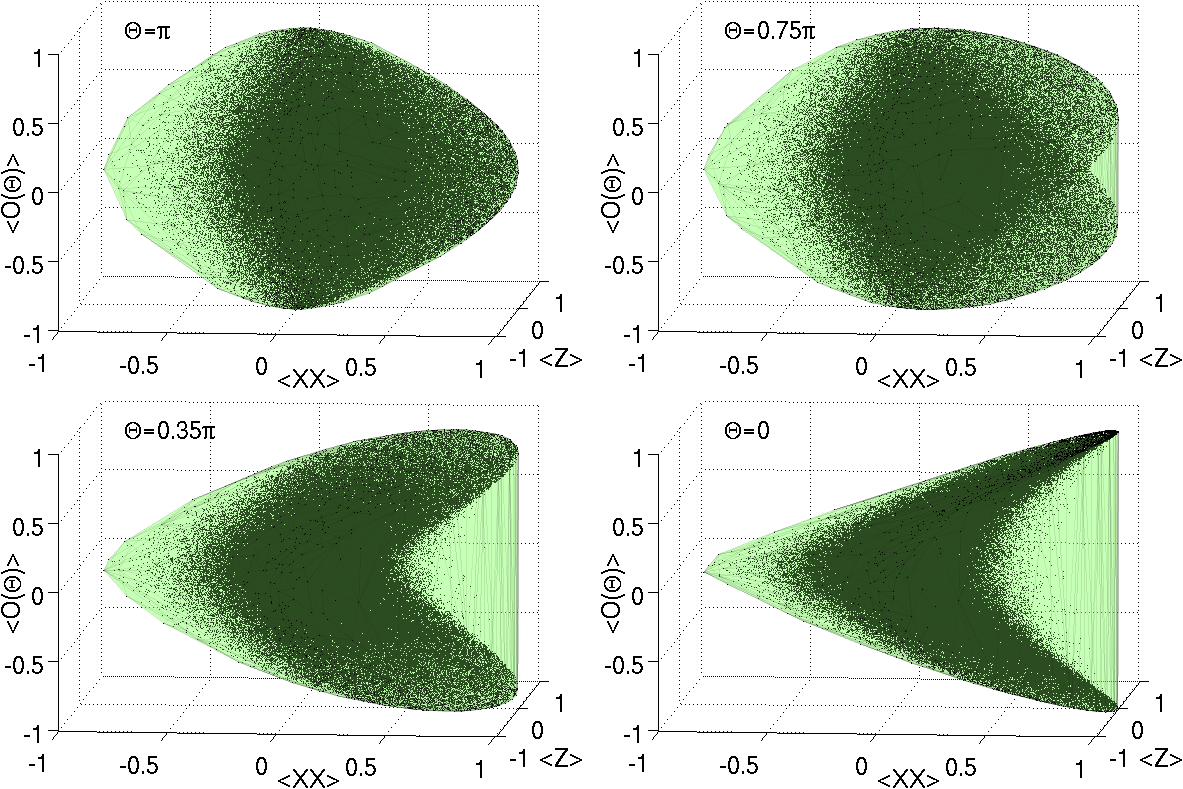}
 \caption{Scatter plot of observables $\braket{XX}$, $\braket{Z}$ and $\braket{O(\Theta)}$ (as defined in Eq. \eqref{eq:randO}) generated from the same set of random states as was also used for \Fig{fig:SFig1}, for $\Theta\in\{0,0.35\pi, 0.75\pi,\pi\}$. We also draw the convex hulls of the respective sets in green. It is clearly visible that the ruled surface is most prominent in the case $\Theta=0$, where $O(0)=X$ corresponds to the true maximum symmetry breaking order parameter associated to the quantum phase transition of \eqref{eq:SIsing0}.}
 \label{fig:SFig2}
\end{figure*}

\begin{figure}[t]
 \centering
 \includegraphics[width=\linewidth,keepaspectratio=true]{\figpath/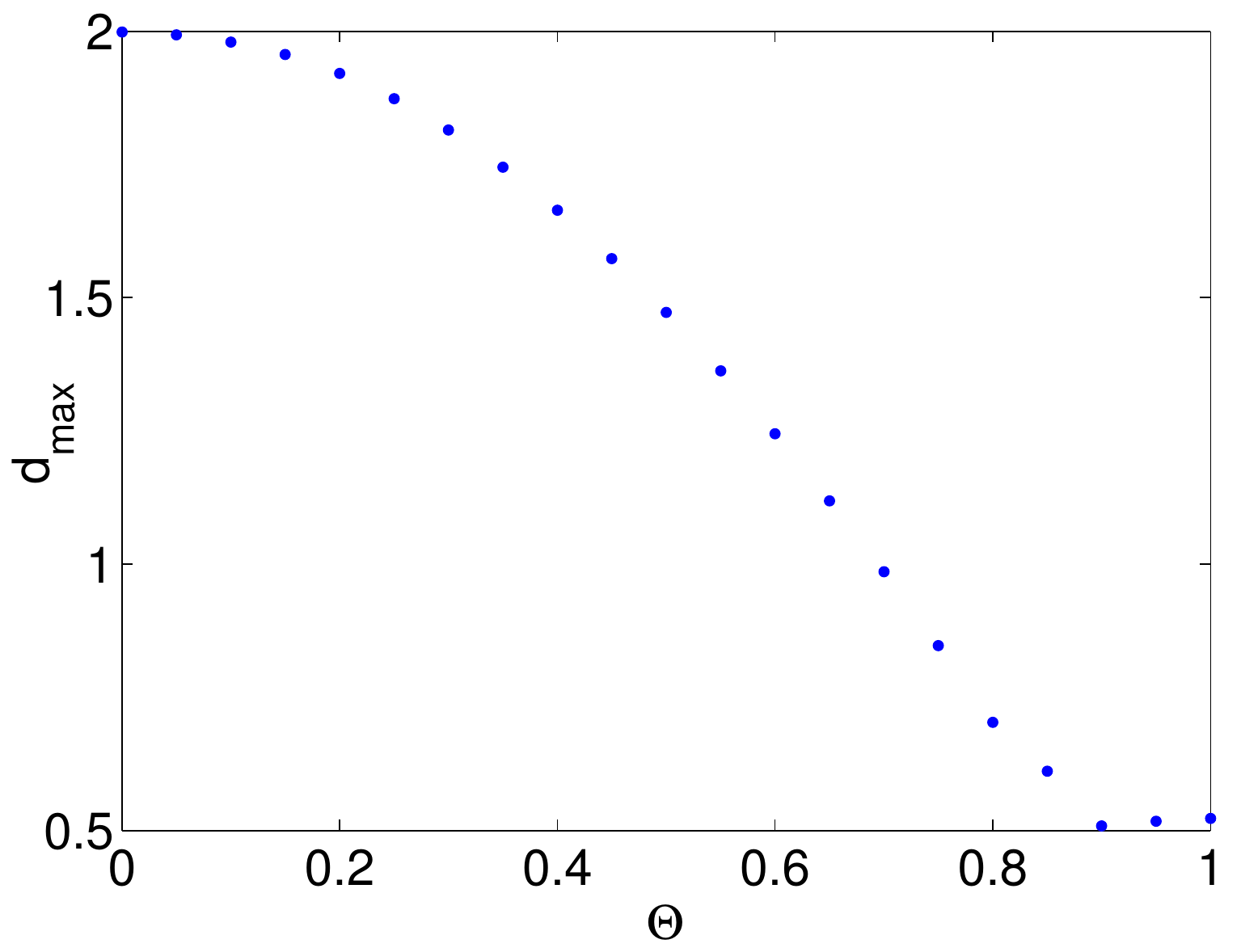}
 \caption{Maximum distance $d_{\rm max}$ of neighboring points on the convex hull as defined in \eqref{eq:dmax}, vs. angle $\Theta$. This quantity clearly takes its maximum value of $d_{\rm max}=2$ at $\Theta=0$, where $O(0)=X$.}
 \label{fig:SFig3}
\end{figure}

\section{Order Parameter Optimization}
\label{s:orderparam}
It is argued in the main text that in the case where the order parameter with maximal symmetry breaking is not known, one can still use a random observable, which will in general have a finite overlap with the true order parameter. Using this observable as an additional axis will thus yield a convex body showing a ruled surface, albeit not with maximal extent. One can then optimize over all possible observables of that type to find the observable which maximizes the extent of the ruled surface and thus corresponds to the order parameter with maximal symmetry breaking.

We demonstrate this procedure in the case of the quantum Ising model
\begin{equation}
\mathcal{H}=-J\sum_{\braket{i,j}}X_iX_j - B_{z}\sum_i Z_i.
\label{eq:SIsing0}
\end{equation}
We assume the order parameter to be a local one-site observable and thus optimize over all possible linear combinations of Pauli-matrices $X$, $Y$ and $Z$, with fixed spectral radius \footnote{If the order parameter is not a one-site observable one would have to optimize over multi-site observables, i.e. over linear combinations of products of Pauli-matrices in the case of spin-1/2 systems. Along that line it would be very interesting to know if it is possible to a priori determine (possibly by different means) just from a random scatter plot if there is a symmetry breaking order parameter at all.}

As the observable $Z$ however is already present in the Hamiltonian itself, it is enough to consider only linear combinations of $X$ and $Y$. We thus define the one-parameter family of observables
\begin{equation}
 O(\Theta)=\cos(\Theta/2)\,X + \sin(\Theta/2)\,Y,\quad 0\leq\Theta<\pi.
 \label{eq:randO}
\end{equation} 
and draw convex sets from random states, using $\braket{O(\Theta)}$ as a third axis, and vary $\Theta$. \Fig{fig:SFig2} shows instances of these sets, together with their convex hulls, for certain selected values of $\Theta$, all generated from the same collection of random states that was also used to generate \Fig{fig:SFig1}. It is clearly visible that the ruled surface in the form of a distinct lobe structure is most prominent for $O(\Theta=0)=X$, whereas there is absolutely no signature of a ruled surface for $O(\Theta=\pi)=Y$.

To quantify the extent of the ruled surface emerging in these plots we 
calculate the maximum distance
\begin{equation}
 d_{\rm max}=\max_{\braket{ij}}|x_{i}-x_{j}|,
 \label{eq:dmax}
\end{equation} 
where $x_{i}$ and $x_{j}$ are neighboring points on the convex hull of the set. This quantity is naturally maximized when the optimal order parameter with maximum symmetry breaking is used as a third axis and is plotted against $\Theta$ for the present case in \Fig{fig:SFig3}, where $d_{\rm max}$ shows a definite maximum at $\Theta=0$.

As visual characterization in terms of ruled surfaces becomes especially hard for convex sets in more than three dimensions (i.e. with more than three observables), \eqref{eq:dmax} can serve as a good quantity to detect ruled surfaces also in these cases.

\bibliography{Convex}

\end{document}